\documentclass[amssymb,useAMS,prd,aps,amsmath,superscriptaddress,nofootinbib,twocolumn,preprintnumbers]{revtex4}
\pdfoutput=1

\usepackage{color}
\usepackage{pslatex}
\usepackage{pdfpages}
\usepackage{graphicx}
\usepackage{psfrag}
\usepackage{pdftricks}
\usepackage{multirow}
\usepackage{graphicx}
\usepackage[none]{hyphenat}
\usepackage{url}
\usepackage[breaklinks, colorlinks, citecolor=blue]{hyperref}
\usepackage{ifthen}
\usepackage{ulem}
\def\eprinttmp@#1arXiv:#2 [#3]#4@{
\ifthenelse{\equal{#3}{x}}{\href{http://arxiv.org/abs/#1}{#1}}{\href{http://arxiv.org/abs/#2}{arXiv:#2} [#3]}}
\providecommand{\eprint}[1]{\eprinttmp@#1arXiv: [x]@}
\newcommand{\adsurl}[1]{\href{#1}{ADS}}

\begin{document}

\title{Constraining Dark Matter--Neutrino Interactions using the CMB and Large-Scale Structure}

\author{Ryan J. Wilkinson}
\affiliation{Institute for Particle Physics Phenomenology, Durham University, Durham, DH1 3LE, United Kingdom}

\author{C\'eline B\oe hm}
\affiliation{Institute for Particle Physics Phenomenology, Durham University, Durham, DH1 3LE, United Kingdom}
\affiliation{LAPTH, U. de Savoie, CNRS,  BP 110, 74941 Annecy-Le-Vieux, France}

\author{Julien Lesgourgues}
\affiliation{LAPTH, U. de Savoie, CNRS,  BP 110, 74941 Annecy-Le-Vieux, France}
\affiliation{Institut de Th\'eorie des Ph\'enom\`enes Physiques, \'Ecole Polytechnique F\'ed\'erale de Lausanne, CH-1015, Lausanne, Switzerland}
\affiliation{CERN, Theory Division, CH-1211 Geneva 23, Switzerland}


\begin{abstract}
We present a new study on the elastic scattering cross section of dark matter (DM) and neutrinos using the latest cosmological data from Planck and large-scale structure experiments. We find that the strongest constraints are set by the Lyman-$\alpha$ forest, giving $\sigma_{\rm{DM}-\nu} \lesssim 10^{-33} \left(m_{\rm{DM}}/\rm{GeV}\right) \ \rm{cm^2}$ if the cross section is constant and a present-day value of $\sigma_{\rm{DM}-\nu} \lesssim 10^{-45} \left(m_{\rm{DM}}/\rm{GeV}\right) \ \rm{cm^2}$ if it scales as the temperature squared. These are the most robust limits on DM--neutrino interactions to date, demonstrating that one can use the distribution of matter in the Universe to probe dark (``invisible") interactions. Additionally, we show that scenarios involving thermal MeV DM and a constant elastic scattering cross section naturally predict (i) a cut-off in the matter power spectrum at the Lyman-$\alpha$ scale, (ii) $N_{\rm eff} \sim 3.5 \pm 0.4$, (iii) $H_0 \sim 71 \pm 3 ~{\mathrm{km}}~{\mathrm{s}}^{-1}~{\mathrm{Mpc}}^{-1}$ and (iv) the possible generation of neutrino masses.

\end{abstract}
\preprint{IPPP/14/03, DCPT/14/06, CERN-PH-TH-2014-013, LAPTH-006/14}

\date{\today}

\maketitle


\section{Introduction}
\label{sec:intro}

Over the last few decades, it has become clear that a large fraction of the Universe is in the form of an invisible material known as dark matter (DM). Recent Cosmic Microwave Background (CMB) results~\cite{Hinshaw:2012aka,Hou:2012xq,Sievers:2013ica,Ade:2013ktc} strongly support the existence of DM, but its nature remains a mystery. The general assumption is that DM consists of cold, massive particles (CDM). However, recent work has shown that small couplings with Standard Model particles (in particular, neutrinos~\cite{Boehm:2000gq,Boehm:2003xr,Boehm:2004th,Mangano:2006mp,Serra:2009uu}, photons~\cite{Boehm:2001hm,Sigurdson:2004zp,Wilkinson:2013kia,Dolgov:2013una} and baryons~\cite{Chen:2002yh,Dvorkin:2013cea,Aviles:2011ak}) cannot yet be ruled out using cosmological data alone and are indeed expected in several extensions of the Standard Model e.g. Refs.~\cite{Boehm:2003hm,Boehm:2006mi,Farzan:2010mr,Lindner:2011it,Law:2013saa} . It is also possible that DM interacts with other putative particles in the dark sector (see e.g. Refs. ~\cite{Ackerman:2008gi,Cyr-Racine:2013fsa,Andreas:2013iba,Franca:2013zxa}) but we will not consider this case here.

Interactions of DM beyond gravity lead to a suppression of the primordial density fluctuations, erasing structures with a size smaller than the ``collisional damping scale"~\cite{Boehm:2000gq,Boehm:2004th}. This produces noticeable signatures in the CMB and matter power spectrum, and ultimately impacts on the large-scale structure (LSS) of the Universe we observe today. The effect is enhanced if DM scatters off relativistic particles e.g. neutrinos and photons in the radiation-dominated era, allowing one to set competitive limits on these interactions in the early Universe.

Unlike direct~\cite{Ahmed:2010wy,Aprile:2012nq,Akerib:2013tjd} and indirect~\cite{Adriani:2008zr,Abbasi:2009uz,Ackermann:2012qk} detection experiments, the results obtained from such analyses are model-independent. Furthermore, any theory that predicts interactions between DM and the visible sector must satisfy these constraints. In this work, we focus on DM--neutrino interactions (a similar study for DM--photon interactions can be found in Ref.~\cite{Wilkinson:2013kia}). We use the latest CMB data from the Planck satellite~\cite{Ade:2013zuv} and observations of large-scale structure from the Lyman-$\alpha$ forest~\cite{Viel:2013fqw} to both update and improve the previous results of Refs.~\cite{Boehm:2000gq,Boehm:2003xr,Boehm:2004th,Mangano:2006mp,Serra:2009uu}.

The paper is organised as follows. In Sec.~\ref{sec:imp}, we recall the modified Euler equations that we use to incorporate DM--neutrino interactions and describe their implementation in the Boltzmann code {\sc class}\footnote{\tt class-code.net}~\cite{Lesgourgues:2011re,Blas:2011rf}. In Sec.~\ref{sec:results}, we present our bounds on the scattering cross section from the CMB angular power spectrum (Sec.~\ref{subsec:cmb}) and the LSS matter power spectrum (Sec.~\ref{subsec:lss}). The significance of our results for specific DM models is discussed in Sec.~\ref{sec:disc} and conclusions are provided in Sec.~\ref{sec:conc}.

\section{DM--Neutrino interactions}
\label{sec:imp}

The modified Euler equations for DM--neutrino interactions can be written as\footnote{For simplicity, we use the Newtonian gauge, assuming a flat Universe and taking derivatives with respect to conformal time. Our notation is consistent with Ref.~\cite{Ma:1995ey}.}~\cite{Boehm:2001hm,Mangano:2006mp}:
\begin{eqnarray}
\label{euler}
\nonumber
\dot \theta_\nu
 & = &   k^2 \psi + k^2 \left(\frac{1}{4} \delta_\nu - \sigma_\nu\right) - \dot \mu (\theta_\nu - \theta_{\rm DM})~, \\
\dot \theta_{\rm{DM}}
 & = &   k ^2\psi - {\cal H} \theta_{\rm{DM}} - S^{- 1} \dot \mu (\theta_{\rm{DM}} - \theta_\nu)~,
\end{eqnarray}
where $\theta_\nu$ and $\theta_{\rm{DM}}$ are the neutrino and DM velocity divergences, $k$ is the comoving wavenumber, $\psi$ is the gravitational potential, $\delta_\nu$ and $\sigma_\nu$ are the neutrino density fluctuation and anisotropic stress potential, and ${\cal H}=(\dot a / a)$ is the conformal Hubble parameter.

The DM--neutrino interaction rate is given by $\dot{\mu} \equiv a\hspace{0.3ex}\sigma_{\rm{DM}-\nu}\hspace{0.3ex}c\hspace{0.3ex}n_{\rm{DM}}$, where $\sigma_{\rm{DM}-\nu}$ is the elastic scattering cross section, $n_{\rm{DM}} = \rho_{\rm{DM}} / m_{\rm{DM}}$ is the DM number density, $\rho_{\rm{DM}}$ is the DM energy density and $m_{\rm{DM}}$ is the DM mass. The factor $S \equiv (3/4)(\rho_{\rm{DM}}/\rho_\nu)$ ensures energy conservation and accounts for the momentum transfer in the elastic scattering process. The new interaction rate is also added to the hierarchy of Boltzmann equations for neutrino temperature and polarisation (in analogy to Thomson scattering terms in the photon Boltzmann hierarchy)\footnote{All necessary modifications are confined to the thermodynamics and perturbation modules of {\sc class} (version 1.7).}.

In most cases, the scattering cross section between DM and neutrinos, $\sigma_{\rm{DM}-\nu}$, will have one of two distinct behaviours: either constant (like Thomson scattering) or proportional to the temperature squared (in analogy to neutrino--electron scattering). This will depend on the particle physics model that is being considered (see Ref.~\cite{Boehm:thesis} for specific examples).

To quantify the effect of DM--neutrino interactions on the evolution of primordial density fluctuations, we introduce the dimensionless quantity
\begin{equation}
\label{u_nudm}
u \equiv \left[\frac{\sigma_{\rm{DM}-\nu}}{\sigma_{\rm Th}} \right] \left[\frac{m_{\rm{DM}}}{100~\rm{GeV}} \right]^{- 1}~,
\end{equation}
where $\sigma_{\rm Th}$ is the Thomson cross section.

Since the magnitude of the $u$ parameter determines the collisional damping scale~\cite{Boehm:2001hm}, the efficiency of small-scale suppression is essentially governed by the ratio of the interaction cross section to the DM mass. For temperature-dependent cross sections, we can write $u = u_0~a^{-2}$, where $u_0$ is the present-day value and $a$ is the cosmological scale factor (normalised to unity today).

\section{Results}
\label{sec:results}

In this section, we present our constraints on the DM--neutrino elastic scattering cross section from the CMB angular power spectrum (Sec.~\ref{subsec:cmb}) and LSS matter power spectrum (Sec.~\ref{subsec:lss}) using the modified version of {\sc class} described above.

\subsection{Cosmic Microwave Background}
\label{subsec:cmb}

\begin{figure}[ht!]
\centering
\hspace{-3ex}
\includegraphics[width=\columnwidth, trim = 3cm 2.6cm 0.9cm 1.8cm]{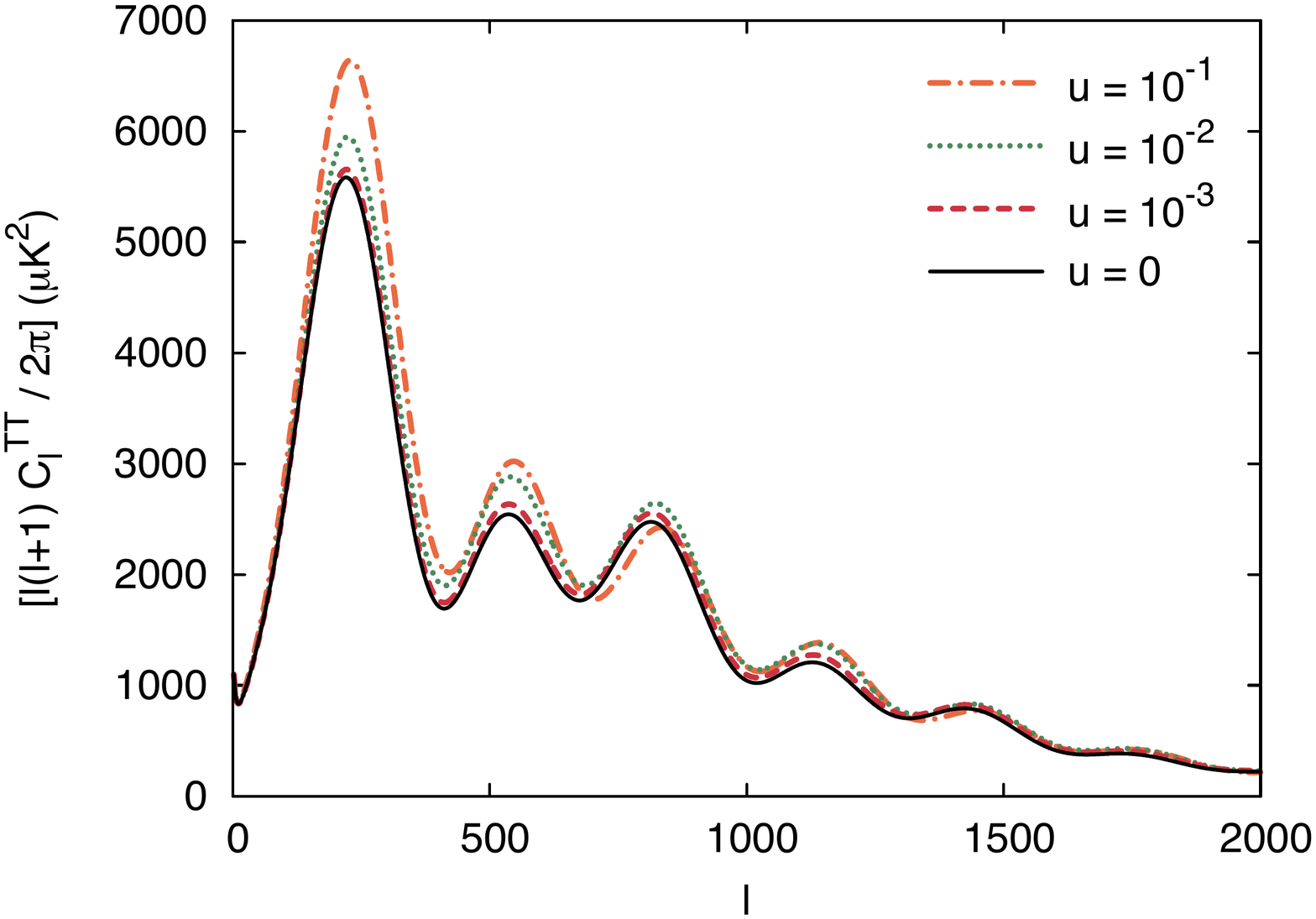}
\vspace{-2ex}
\includegraphics[width=\columnwidth, trim = 3cm 1.8cm 0.9cm 1cm]{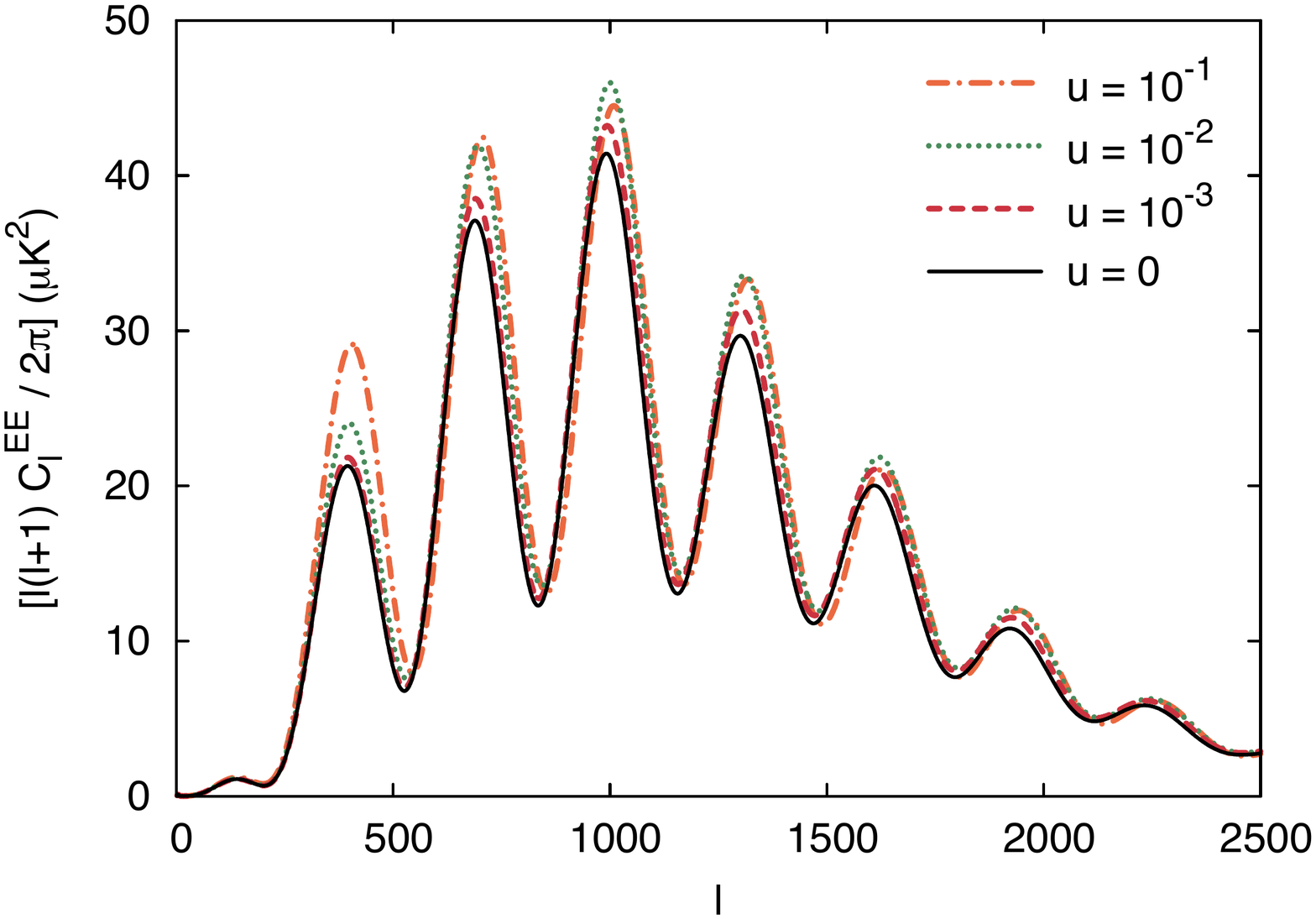}
\vspace{-2ex}
\includegraphics[width=\columnwidth, trim = 3cm 1.1cm 0.9cm 1cm]{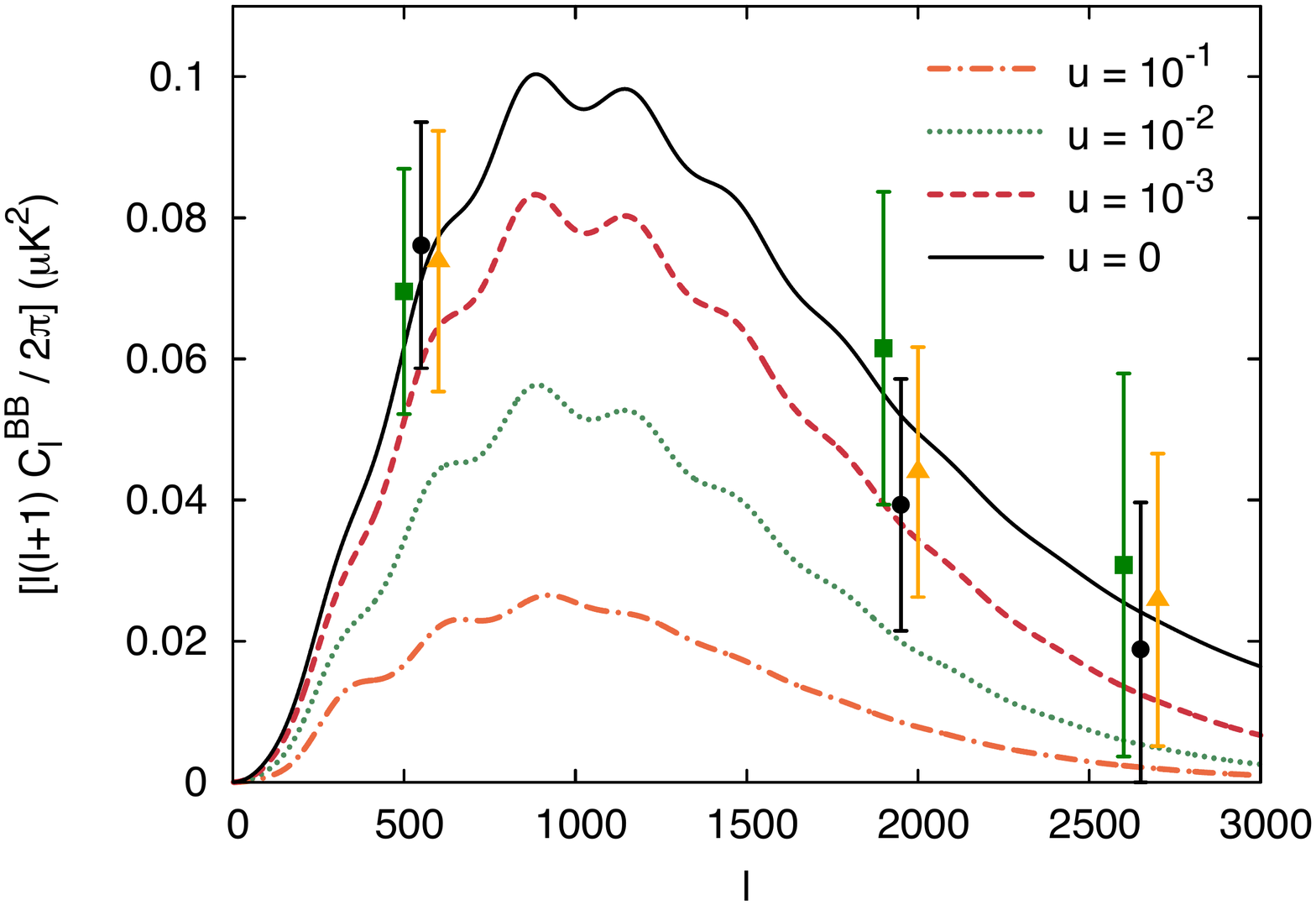}
\vspace{-3ex}
\caption{The effect of DM--neutrino interactions on the $TT$ (top), $EE$ (middle) and $BB$ (bottom) components of the angular power spectrum, where $u \equiv 
         \left[{\sigma_{\rm{DM}-\nu}}/{\sigma_{\mathrm{Th}}} \right]
         \left[{m_{\rm{DM}}}/{100~\rm{GeV}} \right]^{- 1}$ (such that $u = 0$ corresponds to no coupling). We take $\sigma_{\rm{DM}-\nu}$ to be constant and use the `{\it Planck} + WP' best-fit parameters from Ref.~\cite{Ade:2013zuv}. The data points in the $BB$ spectrum are recent measurements from the SPTpol experiment~\cite{Austermann:2012ga}, where the three datasets correspond to $({\hat{\rm{E}}}^{150}{\hat{\phi}}^{\rm{CIB}}) \times {\hat{\rm{B}}}^{150}$, $({\hat{\rm{E}}}^{95}{\hat{\phi}}^{\rm{CIB}}) \times {\hat{\rm{B}}}^{150}$ and $({\hat{\rm{E}}}^{150}{\hat{\phi}}^{\rm{CIB}}) \times {\hat{\rm{B}}}_{\chi}^{150}$ respectively in Ref.~\cite{Hanson:2013hsb}. The new coupling enhances the peaks in the $TT$ and $EE$ spectra, while significantly damping the $B$-modes.\vspace{-5ex}}
\label{fig:c_l}
\end{figure}

The impact of DM--neutrino interactions on the CMB angular power spectrum is illustrated in Fig.~\ref{fig:c_l} for specific values of the parameter $u \equiv \left[{\sigma_{\rm{DM}-\nu}}/{\sigma_{\mathrm{Th}}} \right] \left[{m_{\rm{DM}}}/{100~\rm{GeV}} \right]^{- 1}$. We consider a flat $\Lambda$CDM model (with the only addition being the DM--neutrino coupling), where the cosmological parameters are taken from the one-year data release of Planck~\cite{Ade:2013zuv}. 
We show the impact of a constant cross section in Fig.~\ref{fig:c_l}, however, the effects are similar for temperature-dependent cross sections.

In the $TT$ (top panel) and $EE$ (middle panel) components of the CMB spectrum, we see an increase in the magnitude of the Doppler peaks and a slight shift to larger $l$ with respect to vanilla $\Lambda$CDM ($u = 0$), which can be understood as follows:

The shape of the CMB spectrum is affected by the gravitational force felt by the coupled photon--baryon fluid before decoupling. In principle, this force receives contributions from the distribution of free-streaming neutrinos and from that of slowly-clustering DM. In fact, when decomposing the solution to the system of cosmological perturbations into slow modes and fast modes~\cite{steven2008cosmology,Voruz:2013vqa}, one sees that the photon--baryon and neutrino perturbations are described by fast modes, while the DM perturbations are described by slow modes. This implies that the photon--baryon fluid only has significant gravitational interactions with the free-streaming neutrinos.

This interaction is especially important during radiation domination and soon after Hubble crossing, when the photon--baryon perturbation receives a gravitational boost. This boost is attenuated by the fact that neutrinos free-stream, develop anisotropic stress and cluster less efficiently then e.g. a relativistic perfect fluid. Modes crossing the Hubble radius during matter domination do not experience this effect because the gravitational potential is then constant, while DM perturbations grow in proportion to the scale factor.

In the presence of an efficient DM--neutrino interaction term, DM experiences damped oscillations like neutrinos instead of slow gravitational clustering~\cite{Mangano:2006mp}. Thus, DM perturbations also contribute to the fast modes. At the same time, neutrinos are bound to DM particles and do not free-stream; their anisotropic stress is reduced, making them behave more like a relativistic perfect fluid~\cite{Serra:2009uu}. Both effects contribute to the patterns seen in Fig.~\ref{fig:c_l}: 
\begin{enumerate}
\item
When perturbations cross the Hubble radius during radiation domination, the photon--baryon fluid feels the gravitational force from neutrinos with reduced anisotropic stress and stronger clustering; this increases the gravitational boost effect. This mechanism can potentially enhance all peaks but the first one, although the scale at which this effect is important depends on the time at which neutrinos decouple from DM. 
\item
As long as DM and neutrinos are tightly coupled, the sound speed in this effective fluid is given by $c^2_{\rm{DM}-\nu} = {[3(1+3 \rho_\mathrm{DM}/4 \rho_\nu)]}^{-1}$, instead of $c^2_{{\rm b}-\gamma} = {[3(1+3 \rho_{\rm b}/4 \rho_\gamma)]}^{-1}$ in the baryon--photon fluid. The ratio $\rho_\mathrm{DM}/\rho_\nu$ is always larger than the ratio $\rho_{\rm b}/\rho_\gamma$ so the DM--neutrino fluid has a smaller sound speed. Through gravitational interactions and a ``DM--neutrino drag'' effect, the wavelength of the baryon--photon sound waves is then slightly reduced and the acoustic peaks in the temperature and polarisation spectra appear at slightly larger $l$.
\item
When perturbations cross the Hubble radius during matter domination, if DM is still efficiently coupled to neutrinos, it contributes to the fast mode solution. Thus, DM is gravitationally coupled to the photon--baryon fluid, leading to a gravitational boosting effect (unlike in the standard model for which metric fluctuations are frozen during matter domination). This effect contributes to the enhancement of the first peak.
\item 
In the temperature spectrum, there is a well-known asymmetry between the amplitude of the first odd and even peaks, due to the fact that oscillations in the effective temperature $(\delta T/T + \psi)$ (where $\psi$ is one of the two metric perturbations in the Newtonian gauge) are centred around the mean value $\langle \delta T/T + \psi \rangle \sim - (3 \rho_{\rm b}/4\rho_\gamma) \psi$. If DM is still efficiently coupled to neutrinos at the time of photon decoupling, the metric fluctuations are strongly suppressed, and the oscillations are centred on zero. This has the opposite effect to increasing the baryon density; it slightly enhances even peaks and suppresses odd peaks.
\item
Finally, if DM is still efficiently coupled to neutrinos at the time of photon decoupling, the first peak is further enhanced by a stronger early integrated Sachs-Wolfe effect. This takes place after photon decoupling as a consequence of the fact that metric fluctuations vary with time as long as DM remains efficiently coupled to neutrinos.
\end{enumerate}

Note that among all these effects, the first two can occur even for a small DM--neutrino cross section, since they only assume that neutrinos are coupled to DM until some time near the end of radiation domination. The last three effects are only present for very large cross sections, such that DM is still coupled to neutrinos at the beginning of matter domination. All five effects can be observed in Fig.~\ref{fig:c_l} for $u = 10^{-3}$ or larger (corresponding to $\sigma_{\rm{DM}-\nu} \gtrsim 10^{-29} \left(m_{\rm{DM}}/\rm{GeV}\right) \ \rm{cm^2}$). However, we will see in Sec.~\ref{subsec:lss} that these values are not compatible with Lyman-$\alpha$ data; for realistic cross sections, the only effect on the CMB spectrum is a small enhancement and shifting of the high-$l$ peaks.

To efficiently sample the parameter space and account for any degeneracies, we ran the Markov Chain Monte Carlo code {\sc Monte Python}\footnote{\tt montepython.net}~\cite{Audren:2012wb} combined with the one-year data release from Planck, provided by the Planck Legacy Archive\footnote{\tt  pla.esac.esa.int/pla/aio/planckProducts.html}~\cite{Ade:2013ktc}. In particular, we used the high-$\ell$ and low-$\ell$ temperature data of Planck combined with the low-$\ell$ WMAP polarisation data (corresponding to `{\it Planck} + WP' in Ref.~\cite{Ade:2013zuv}).

We varied the parameters of the minimal flat $\Lambda$CDM cosmology\footnote{Our base model is the same as in the Planck analysis, except for one detail: we use the approximation of massless neutrinos, while the Planck collaboration always assumes one massive neutrino species with a mass of 0.06~eV~\cite{Ade:2013zuv}. We chose the massless option simply to speed up computations, however, it has very little impact. At the level of precision of Planck, such a small neutrino mass only affects the CMB through a slight shift in the angular diameter distance. This can be exactly compensated by a decrease in the Hubble parameter of about $\Delta h \simeq -0.1 (m_\nu / 1\, \mathrm{eV})$~\cite{Ade:2013zuv}. Therefore, had we adopted the same base model as in the Planck papers, we would obtain a best-fit value of $100~h$ that is smaller by $\sim 0.6$. However, the other results (i.e. the uncertainty on $h$, the mean values and uncertainties of the other parameters, and the maximum likelihood value) would be unchanged. \label{mnu}}, namely: the baryon density ($\Omega_{\rm{b}}h^2$), the dark matter density ($\Omega_{\rm{DM}}h^2$), the reduced Hubble parameter ($h$), the primordial spectrum amplitude ($A_s$), the scalar spectral index ($n_s$) and the redshift of reionisation ($z_{\rm{reio}}$), supplemented by the additional parameter $u \equiv \left[{\sigma_{\rm{DM}-\nu}}/{\sigma_{\mathrm{Th}}} \right]\left[{m_{\rm{DM}}}/{100~\rm{GeV}} \right]^{- 1}$. In a second run, we also allowed the effective number of neutrino species, $N_{\rm eff}$, to vary from the standard value of 3.046~\cite{Mangano:2005cc}. Finally, we marginalised over the nuisance parameters listed in Ref.~\cite{Ade:2013zuv}.

\begin{figure*}
\centering
\includegraphics[scale=0.44]{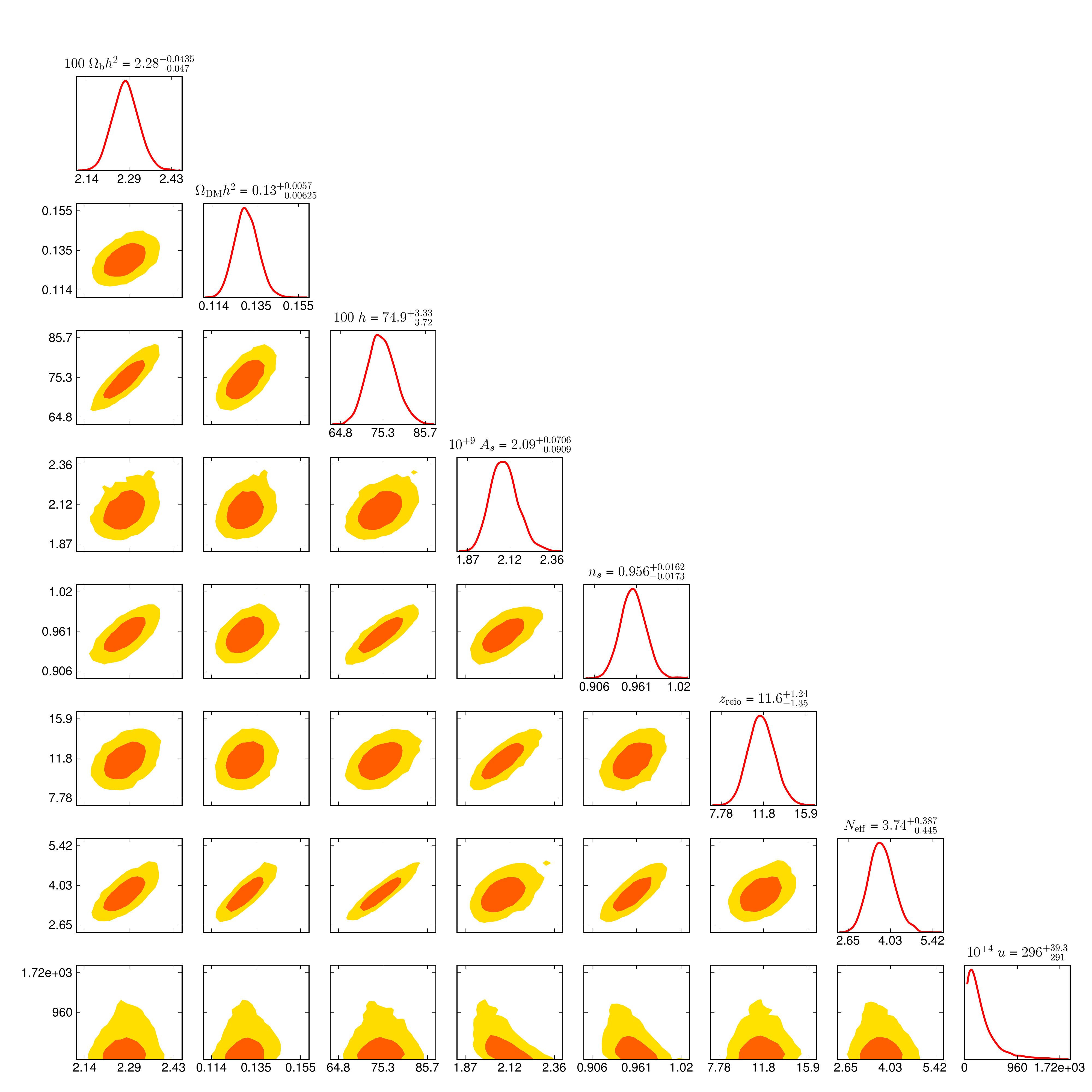}
\caption{Triangle plot showing the one and two-dimensional posterior distributions of the cosmological parameters set by Planck for a constant cross section, with $u$ and $N_{\rm eff}$ as free parameters. The contours correspond to 68\% and 95\% CL.}
\label{fig:planck_fit1}
\end{figure*}

\begin{figure*}
\centering
\includegraphics[scale=0.44]{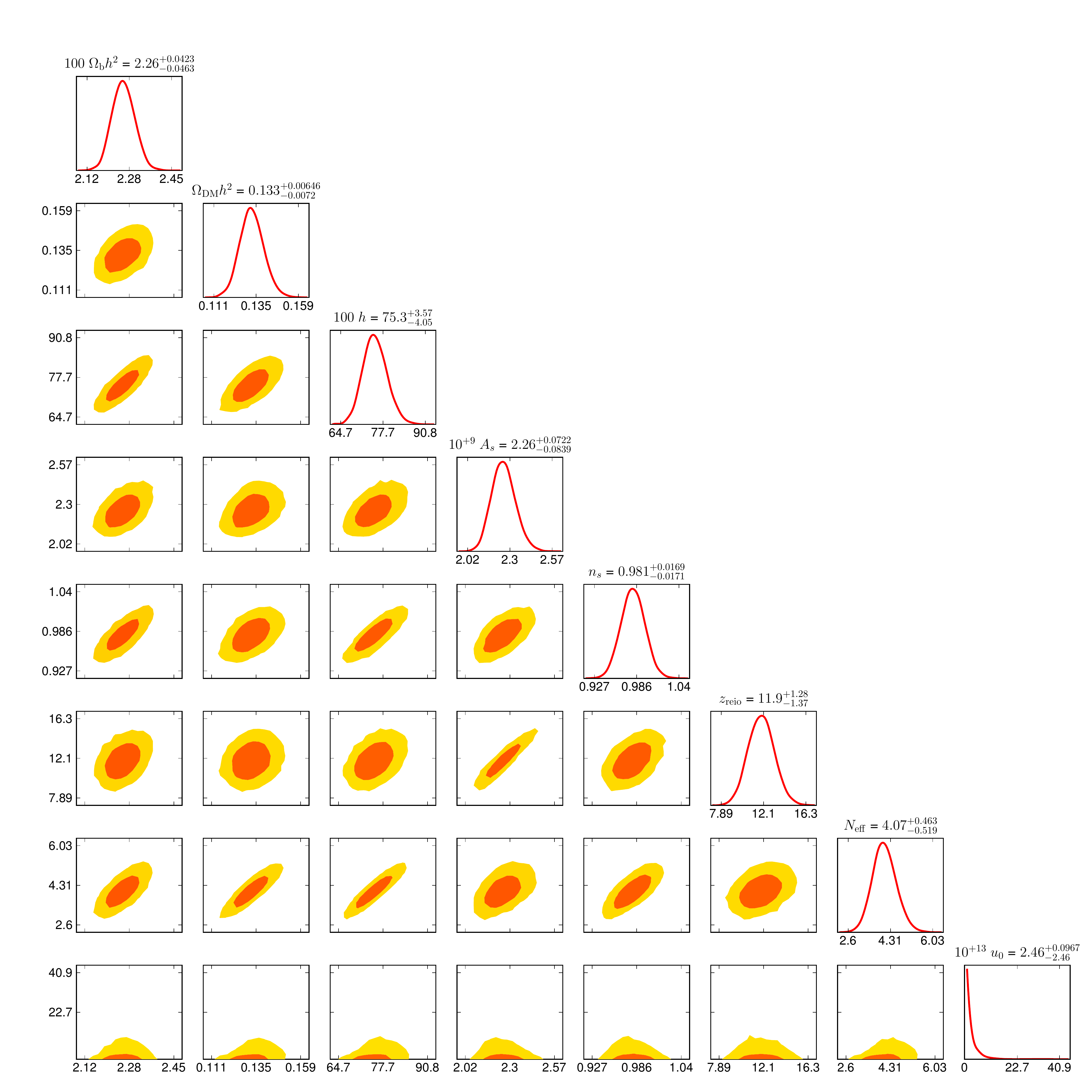}
\caption{Triangle plot showing the one and two-dimensional posterior distributions of the cosmological parameters set by Planck for a temperature-dependent cross section, with $u$ and $N_{\rm eff}$ as free parameters. The contours correspond to 68\% and 95\% CL.}
\label{fig:planck_fit2}
\end{figure*}

The bounds on the various cosmological parameters are listed in Table~\ref{tab:params1}, and illustrated in Figs.~\ref{fig:planck_fit1} and~\ref{fig:planck_fit2} for constant and temperature-dependent cross sections respectively (where we omit the nuisance parameters for clarity).

\begin{table*}
  \begin{tabular*}{0.98\textwidth}{@{\extracolsep{\fill} }c|ccccccccc}
     \hline
~~~~ &   $100~\Omega_{\rm{b}} h^2$ & $\Omega_{\rm{DM}} h^2$  & $100~h$ &
     $10^{+9}~A_{s }$ & $n_s$ & $z_{\rm{reio}}$ & $N_{\rm eff}$ & $10^{+2}~u$  & $10^{+13}~u_0$  \\ \hline 
     \vspace{-2ex}
     &&&&&&&&\\
      ~~No interaction~~ & $2.205_{-0.028}^{+0.028}$ & $0.1199_{-0.0027}^{+0.0027}$ &
     $67.3_{-1.2}^{+1.2}$ & $2.196_{-0.060}^{+0.051}$ &
     $0.9603_{-0.0073}^{+0.0073}$ & $11.1_{-1.1}^{+1.1}$ & $(3.046)$ & $-$ & $-$ \\ [1.5ex]
     \vspace{-2ex}
     &&&&&&&&\\
      ~~~~ & $2.238_{-0.041}^{+0.041}$ & $0.1256_{-0.0055}^{+0.0055}$ &
     $70.7_{-3.2}^{+3.2}$ & $2.251_{-0.085}^{+0.069}$ &
     $0.977_{-0.016}^{+0.016}$ & $11.6_{-1.3}^{+1.3}$ & $3.51_{-0.39}^{+0.39}$ & $-$ & $-$ \\ [1.5ex] \hline
     \vspace{-2ex}
     &&&&&&&\\
     \vspace{-2.5ex}
     &&&&&&&&\\
 ~~$\sigma_{\rm{DM}-\nu}$ constant~~ & $2.225_{-0.033}^{+0.029}$ & $0.1211_{-0.0030}^{+0.0027}$ &
     $69.5_{-1.2}^{+1.2}$ & $2.020_{-0.065}^{+0.063}$ &
     $0.9330_{-0.0095}^{+0.0104}$ & $10.8_{-1.1}^{+1.1}$ & $(3.046)$ & $< 3.99$ & $-$ \\[1.5ex]
     \vspace{-2ex}
     &&&&&&&&\\
 ~~~~ & $2.276_{-0.048}^{+0.043}$ & $0.1299_{-0.0061}^{+0.0059}$ &
     $75.0_{-3.7}^{+3.4}$ & $2.086_{-0.089}^{+0.068}$ &
     $0.956_{-0.016}^{+0.017}$ & $11.6_{-1.3}^{+1.2}$ & $3.75_{-0.43}^{+0.40}$ & $< 3.27$ & $-$
\\[1.5ex]\hline 
     \vspace{-2ex}
     &&&&&&&&\\
 ~~$\sigma_{\rm{DM}-\nu} \propto T^2$~~ & $2.197_{-0.028}^{+0.028}$ & $0.1197_{-0.0027}^{+0.0027}$ &
     $67.8_{-1.2}^{+1.2}$ & $2.167_{-0.059}^{+0.052}$ &
     $0.9527_{-0.0085}^{+0.0086}$ & $10.8_{-1.1}^{+1.1}$ & $(3.046)$ & $-$ & $< 0.54$ \\[1.5ex]
     \vspace{-2ex}
     &&&&&&&&\\
 ~~~~ & $2.262_{-0.046}^{+0.042}$ & $0.1326_{-0.0072}^{+0.0065}$ &
     $75.3_{-4.0}^{+3.6}$ & $2.257_{-0.084}^{+0.072}$ &
     $0.981_{-0.017}^{+0.017}$ & $11.9_{-1.4}^{+1.3}$ & $4.07_{-0.52}^{+0.46}$ & $-$ & $< 2.56$ \\[1.5ex]\hline 
   \end{tabular*}
\caption{Mean values and minimum credible intervals at 68\% CL of the cosmological parameters set by the `{\it Planck} + WP' dataset for (i) no DM--neutrino interaction, (ii) a constant cross section and (iii) a temperature-dependent cross section, where $u \equiv \left[{\sigma_{\rm{DM}-\nu}}/{\sigma_{\mathrm{Th}}} \right] \left[{m_{\rm{DM}}}/{100~\rm{GeV}} \right]^{- 1}$. In each of these models, we consider either a fixed $N_{\mathrm{eff}}$ (first row) or varying $N_\mathrm{eff}$ (second row). The case without an interaction is shown for comparison, using data from Ref.~\cite{Ade:2013zuv} and the Planck Explanatory 
Supplement ({\tt http://www.sciops.esa.int/wikiSI/planckpla/}). For a fair comparison of $h$ values between the interacting and non-interacting scenarios, one should subtract 0.6 from the mean $100~h$ values of the last four lines, for the reason explained in Footnote~\ref{mnu}.}
\label{tab:params1}
\end{table*}

Fixing $N_{\rm eff} = 3.046$, we find that the data prefers an elastic scattering cross section of
\begin{equation}
\sigma_{\rm{DM}-\nu} \leq~3 \times 10^{-28} \left(m_{\rm{DM}}/\rm{GeV}\right) \ \rm{cm^2}~,
\label{eq:bound}
\end{equation}
if it is constant and
\begin{equation}
\sigma_{\rm{DM}-\nu,0} \leq~4 \times 10^{-40} \left(m_{\rm{DM}}/\rm{GeV}\right) \ \rm{cm^2}~,
\end{equation}
for the present-day value if it is proportional to the temperature squared (at 68\% CL).

The bound on the constant cross section is rather weak due to significant degeneracies with the other parameters (in particular: $h$, $A_s$ and $n_s$). By performing additional runs, we found that including constraints on $\sigma_8$ from e.g. Planck SZ clusters~\cite{Ade:2013lmv} and CFHTLens~\cite{Heymans:2013fya} does not help to break the degeneracies. The reason is that for most allowed models, deviations from $\Lambda$CDM occur at scales smaller than those probed by these experiments.

In the standard case of no DM--neutrino interaction, the Planck collaboration found that allowing $N_{\rm eff}$ to vary as a free parameter does not significantly improve the goodness-of-fit for `{\it Planck} + WP' data. However, it has the remarkable property of enlarging the bounds on $h$, in such a way as to relax the tension with direct measurements of the local Hubble expansion (without conflicting with Baryon Acoustic Oscillation data)~\cite{Ade:2013zuv}.

This is a result of a well-known parameter degeneracy, involving at least $N_\mathrm{eff}$, $h$ and $\Omega_{\rm m} h^2$. This degeneracy comes from the fact that by simultaneously enhancing the radiation, matter and cosmological constant densities in the Universe, one does not change the characteristic redshifts and distances affecting the CMB spectrum up to $l\sim 800$. Nevertheless, this direction of degeneracy can be constrained because additional degrees of freedom in $N_{\rm eff}$ lead to a stronger Silk damping effect, which is clearly visible for $l \gtrsim 800$. Thus, the varying $N_{\rm eff}$ model is not preferred by Planck alone, but has the potential to reconcile different cosmological probes that are otherwise in moderate ($\sim$ 2.5$\sigma$) tension.

If we now introduce DM--neutrino interactions, the model with varying $N_{\rm eff}$ turns out to be even more interesting. As in the standard case, it does not significantly improve the goodness-of-fit to `{\it Planck} + WP' data (the effective $\chi^2$ decreases by about two for a constant cross section and 0.5 for a temperature-dependent cross section). However, it opens up an even wider degeneracy in parameter space because the enhancement of the temperature spectrum shown in Fig.~\ref{fig:c_l} can, to some extent, counteract the additional Silk damping caused by a large $N_{\rm eff}$ or $h$.

Therefore, as can be seen in Table~\ref{tab:params1}, with the addition of DM--neutrino interactions, the `{\it Planck} + WP' data can accommodate very large values of $N_{\rm eff}$ (compatible with one thermalised species of extra relics) and $h$ (in excellent agreement with direct measurements at the 1$\sigma$ level \cite{Riess:2011yx,Freedman:2012ny}).

Allowing $N_{\rm eff}$ to vary, we obtain slightly different bounds on the scattering cross section:
\begin{equation}
\sigma_{\rm{DM}-\nu} \leq~2 \times 10^{-28} \left(m_{\rm{DM}}/\rm{GeV}\right) \ \rm{cm^2}~,
\label{eq:bound2}
\end{equation}
if it is constant and
\begin{equation}
\sigma_{\rm{DM}-\nu,0} \leq~2 \times 10^{-39} \left(m_{\rm{DM}}/\rm{GeV}\right) \ \rm{cm^2}~,
\end{equation}
if it is proportional to the temperature squared (at 68\% CL).

Finally, we can use the $BB$ spectrum (bottom panel of Fig.~\ref{fig:c_l}) to constrain the DM--neutrino cross section. The $B$-modes are significantly suppressed due to the effects of collisional damping (see Refs.~\cite{Boehm:2000gq,Boehm:2004th}). Using the first-season data from the SPTpol experiment~\cite{Hanson:2013hsb} (shown by the data points), we can already set conservative limits on the cross section of
\begin{equation}
\sigma_{\rm{DM}-\nu} \lesssim~10^{-27} \left(m_{\rm{DM}}/\rm{GeV}\right) \ \rm{cm^2}~,
\end{equation}
if it is constant and
\begin{equation}
\sigma_{\rm{DM}-\nu,0} \lesssim~10^{-35} \left(m_{\rm{DM}}/\rm{GeV}\right) \ \rm{cm^2}~,
\end{equation}
if it is proportional to the temperature squared.

Forthcoming polarisation data from e.g. Planck~\cite{Ade:2013ktc}, ACTpol~\cite{Niemack:2010wz}, POLARBEAR~\cite{Kermish:2012eh} and SPIDER~\cite{Crill:2008rd} will improve these results and could provide us with a powerful tool to study DM interactions in the future.

\subsection{Large-Scale Structure}
\label{subsec:lss}

\begin{table*}
  \begin{tabular*}{0.95\textwidth}{@{\extracolsep{\fill} }c|ccccccc}
     \hline
&   $100~\Omega_{\rm{b}} h^2$ & $\Omega_{\rm{DM}} h^2$  & $100~h$ &
     $10^{+9}~A_{s }$ & $n_s$ & $z_{\rm{reio}}$ & $N_{\rm eff}$  \\ \hline 
     \vspace{-2ex}
     &&&&&&&\\
 ~~Lyman-$\alpha$ limit ~~ & $2.246_{-0.042}^{+0.039}$ & $0.1253_{-0.0056}^{+0.0053}$ &
     $71.5_{-3.3}^{+3.0}$ & $2.254_{-0.082}^{+0.069}$ &
     $0.979_{-0.016}^{+0.016}$ & $11.7_{-1.3}^{+1.2}$ & $3.52_{-0.40}^{+0.36}$ \\[1.5ex]\hline 
   \end{tabular*}
\caption{Best-fit values and minimum credible intervals at 68\% CL of the cosmological parameters set by the `{\it Planck} + WP' dataset for a constant DM--neutrino elastic scattering cross section, where we impose the maximum allowed value obtained in Sec.~\ref{subsec:lss}, i.e. $\sigma_{\rm{DM}-\nu} \simeq 10^{-33} \left(m_{\rm{DM}}/\rm{GeV}\right) \ \rm{cm^2}$.
}
\label{tab:params3}
\end{table*}

\begin{figure}
\centering
\includegraphics[width=8.8cm, trim = 3cm 2.5cm 1cm 2cm]{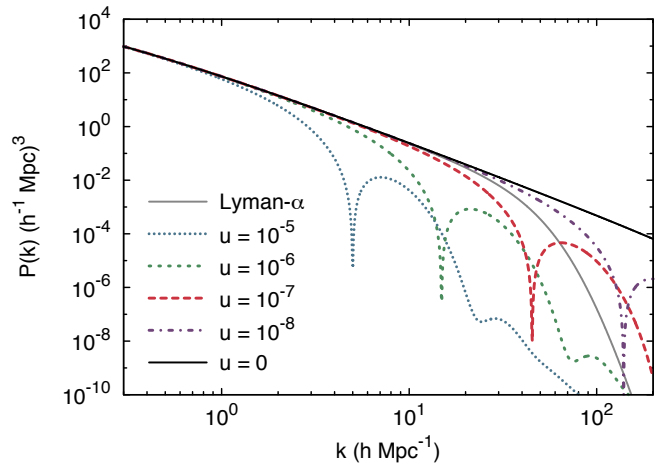}
\caption{The impact of DM--neutrino interactions on the matter power spectrum, where $u \equiv 
         \left[{\sigma_{\rm{DM}-\nu}}/{\sigma_{\mathrm{Th}}} \right]
         \left[{m_{\rm{DM}}}/{100~\rm{GeV}} \right]^{- 1}$ (such that $u = 0$ corresponds to no coupling). We take $\sigma_{\rm{DM}-\nu}$ to be constant and use the `{\it Planck} + WP' best-fit parameters from Ref.~\cite{Ade:2013zuv}. The solid grey curve represents the most recent constraint on warm DM models from the Lyman-$\alpha$ forest~\cite{Viel:2013fqw}. The new coupling produces (power-law) damped oscillations, reducing the number of small-scale structures with respect to vanilla $\Lambda$CDM~\cite{Boehm:2001hm}.}
\vspace{2.5ex}
\label{fig:p(k)}
\end{figure}

The effects of introducing DM--neutrino interactions on the matter power spectrum, $P(k)$, are shown in Fig.~\ref{fig:p(k)} (where for simplicity, we assume that the cross section is constant). We obtain a series of damped oscillations, which suppress power on small scales (see Ref.~\cite{Boehm:2001hm}). For the cross sections of interest, significant damping effects are restricted to the non-linear regime (for which $k \gtrsim 0.2~h~{\rm{Mpc}}^{-1}$).

In general, the reduction of small-scale power for a DM candidate is described by a transfer function, $T(k)$, defined by
\begin{equation}
P(k) = T^2(k)~P_{\rm CDM}(k)~,
\end{equation}
where $P_{\rm CDM}(k)$ is the equivalent matter power spectrum for CDM.

For a non-interacting warm DM (WDM) particle, the transfer function can be approximated by the fitting formula~\cite{Bode:2000gq}:
\begin{equation}
T(k) = [1 + (\alpha k) ^{2 \nu}]^{-5 / \nu}~,
\label{fitmass}
\end{equation}
where
  \begin{equation}
    \alpha = \frac{0.049}{h~{\rm{Mpc}}^{-1}}
    \left(\frac{m_{\rm WDM}}{\rm keV}\right)^{-1.11}
    \left(\frac{\Omega_{\rm DM}}{0.25}\right)^{0.11}
    \left(\frac{h}{0.7}\right)^{1.22},
  \end{equation}
$\nu \simeq 1.12$ and $m_{\rm WDM}$ is the mass of the warm thermal relic~\cite{Viel:2005qj}.

From Fig.~\ref{fig:p(k)}, one can see that cosmological models including DM--neutrino interactions can provide an initial reduction of small-scale power in a similar manner to the exponential cut-off of WDM. The presence of damped oscillations is unimportant for setting limits since we are only interested in the cut-off of the spectrum and the power is already significantly reduced by the first oscillation. However, we note that this difference could allow one to distinguish the two models in high-resolution N-body simulations~\cite{Jascha}.

Using an analysis of the Lyman-$\alpha$ flux from the HIRES~\cite{Vogt:1995zz} and MIKE spectrographs~\cite{2002SPIE.4485..453B}, Ref.~\cite{Viel:2013fqw} obtained a bound on the free-streaming scale of a warm thermal relic, corresponding to a particle mass of $m_{\rm WDM} \simeq 3.3~{\rm keV}$ (or equivalently, $\alpha \simeq 0.012$). This constraint is represented by the solid grey curve in Fig.~\ref{fig:p(k)}.

By comparing models of DM--neutrino interactions with WDM, we can effectively rule out cross sections in which the collisional damping scale is larger than the maximally-allowed WDM free-streaming scale. Taking into account the freedom from the other cosmological parameters, we obtain the conservative upper bounds:
\begin{equation}
\sigma_{\rm{DM}-\nu} \lesssim 10^{-33} \left(m_{\rm{DM}}/\rm{GeV}\right) \ \rm{cm^2}~,
\label{eq:lyman_const}
\end{equation}
if the cross section is constant and
\begin{equation}
\sigma_{\rm{DM}-\nu,0} \lesssim 10^{-45} \left(m_{\rm{DM}}/\rm{GeV}\right) \ \rm{cm^2}~,
\end{equation}
if it scales as the temperature squared.

These limits are significantly stronger than those obtained from the CMB analysis in Sec.~\ref{subsec:cmb} and will improve further with forthcoming data from LSS surveys such as SDSS-III~\cite{Eisenstein:2011sa} and Euclid~\cite{Laureijs:2011gra}. However, CMB constraints are important to compare to as they do not depend on the non-linear evolution of the matter fluctuations.

\begin{figure*}
\centering
\includegraphics[scale=0.44]{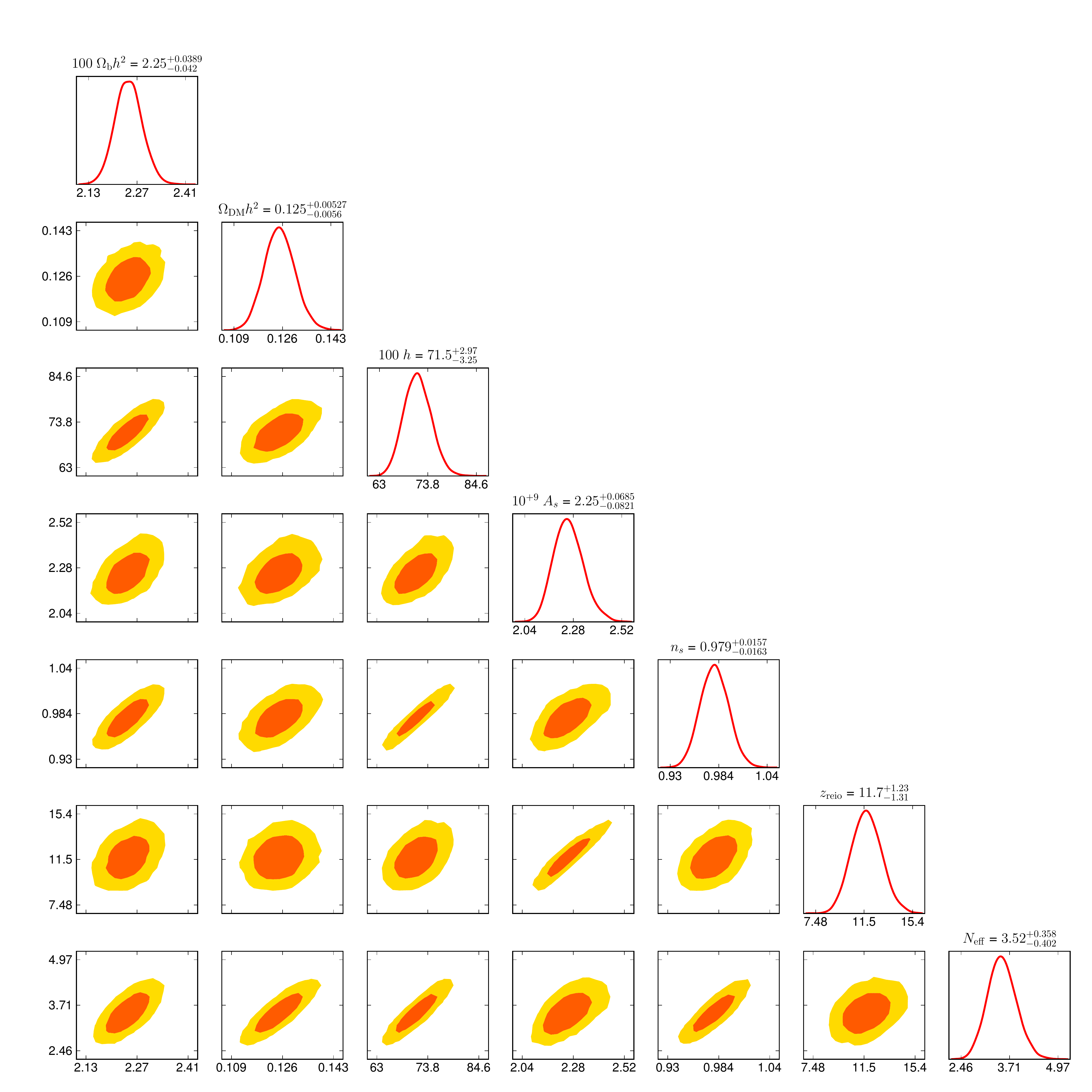}
\caption{Triangle plot showing the one and two-dimensional posterior distributions of the cosmological parameters set by Planck for a constant cross section, where we impose the maximum allowed value obtained in Sec.~\ref{subsec:lss}, i.e. $\sigma_{\rm{DM}-\nu} \simeq 10^{-33} \left(m_{\rm{DM}}/\rm{GeV}\right) \ \rm{cm^2}$. The contours correspond to 68\% and 95\% CL.}
\label{fig:planck_fit3}
\end{figure*}

We can now fix the cross section to be the maximum value allowed by these constraints and redo our CMB analysis. Applying Eq.~\eqref{eq:lyman_const} for a constant cross section, we obtain the bounds on the cosmological parameters shown in Table~\ref{tab:params3} and illustrated in Fig.~\ref{fig:planck_fit3}. These results are similar to the case of no interaction with $N_{\rm eff}$ free to vary, corresponding to the second line in Table~\ref{tab:params1} (especially after correcting the central value of $100~h$ by 0.6, as explained in Footnote~\ref{mnu}). The reason is that the cross section imposed by the Lyman-$\alpha$ data is small enough to not significantly modify the CMB spectrum.

Finally, we note that if more than one species were responsible for the observed DM relic density (which is the case that we consider here), larger values of the elastic scattering cross section would be allowed.

\section{Discussion}
\label{sec:disc}

The results from Section~\ref{sec:results} enable us to constrain DM interactions that cannot be directly probed at the LHC and provide us with direct access to physics beyond the Standard Model in the Early Universe. They are particularly useful for the models proposed in Refs.~\cite{Boehm:2003hm,Boehm:2003xr,Boehm:2006mi} where the DM particle is light ($\sim$ MeV) and interactions with neutrinos can occur through the exchange of a scalar mediator (if DM is fermionic) or a Dirac/Majorana mediator (hereafter referred to as $N$, if DM is a scalar).

Our limits could also be applied to the case of fermionic/scalar DM coupled to a light $U(1)$ gauge boson mediator (referred to as $U$ or $Z'$)~\cite{Boehm:2003hm,Boehm:2006mi} with the caveat that the coupling of such a mediator to neutrinos is constrained by neutrino elastic scattering experiments~\cite{Boehm:2004uq,Chiang:2012ww}.

\subsection{Constant Cross Section}
\label{subsec:const}

In general, one expects the DM--neutrino elastic scattering cross section to be temperature-dependent. However, a constant (i.e. temperature-independent) cross section is predicted either when (i) there is a strong degeneracy between the DM particle and the mediator or (ii) the mediator is extremely light (which, in the case considered here, would imply that DM decays into the mediator plus a neutrino, unless the couplings are very suppressed).

To illustrate point (i), we consider the particular case of a real scalar DM particle coupled to a Majorana mediator, $N$ (an analogue of the sneutrino--neutralino--neutrino coupling in Supersymmetry) in a low effective theory \cite{Boehm:2003hm,Boehm:2006mi}. We then impose a strong mass degeneracy between the DM particle and $N$, i.e. $|m_N - m_{\rm DM}| \lesssim \mathcal{O}(\mathrm{eV})$. In such a scenario, the elastic scattering cross section is expected to be 
\begin{eqnarray}
\sigma_{\rm{DM}-\nu}  &\simeq&  \frac{g^4}{4 \ \pi \ m_{\rm{DM}}^2} \nonumber \\ &\simeq& 3 \times 10^{-33} \ \left(\frac{g}{0.1}\right)^4 \ \left(\frac{m_{\rm DM}}{\rm{GeV}}\right)^{-2} \ \rm{cm^2}~,
\end{eqnarray}
where $g$ is the DM--neutrino coupling.

Applying our Lyman-$\alpha$ constraint from Sec.~\ref{subsec:lss} implies the following relation between the coupling and the DM mass:
\begin{equation}
g \lesssim 0.1 \ \left(\frac{m_{\rm DM}}{\rm{GeV}}\right)^{3/4}~.
\label{eq:g}
\end{equation}

An additional feature of this model is the self-annihilation of DM into neutrinos ($\nu \nu$), with a cross section given by
\begin{equation}
\langle \sigma v \rangle \simeq \frac{g^4}{16 \pi} \ \frac{1}{m_{\rm DM}^2}~\times c~,
\end{equation}
in the primordial Universe~\cite{Boehm:2006mi}. Thus, the annihilation and elastic scattering cross sections are related by
\begin{equation}
\langle \sigma v \rangle  \simeq   \frac{\sigma_{{\rm DM}-\nu}}{4} \ \times c~,
\end{equation}
which gives $\langle \sigma v \rangle \simeq 2 \times 10^{-23}~\left({m_{\rm DM}}/{\rm GeV} \right)~\rm{cm}^3~{\rm s}^{-1}$ if we apply our Lyman-$\alpha$ bound. Conversely, if we impose that the DM annihilation cross section into neutrinos is within the range that is needed to explain the observed DM relic abundance\footnote{The assumption of dominant annihilations into neutrinos at MeV energies makes sense since significant annihilations into charged particles would require new, relatively light (charged) species. Such particles have not been observed, neither directly at the LHC nor in Particle Physics experiments (such as the electron/muon $g-2$~\cite{Boehm:2004gt,Boehm:2007na,Hanneke:2010au,Bouchendira:2010es}).}, we obtain the prediction that
\begin{equation}
\sigma_{{\rm DM}-\nu} \ \simeq \ 4 \times 10^{-36}   \ \left(\frac{\langle \sigma v \rangle}{3 \times 10^{-26}~ \rm{cm}^3~{\rm s}^{-1}}\right)  \ \rm{cm^2}~,
\label{eq:annihilation}
\end{equation}
which is similar to our Lyman-$\alpha$ bound for MeV DM.

Therefore, we deduce that a viable model of MeV DM with a coupling to neutrinos must predict a cut-off in the matter power spectrum at the Lyman-$\alpha$ scale. Note that, in principle, we should also allow for co-annihilations~\cite{Binetruy:1983jf,Griest:1990kh} since we assume a strong mass degeneracy between the DM particle and the mediator. A self-annihilation cross section that is $\sim$ 4 times smaller than the value quoted above would thus give rise to the observed DM abundance.

Interestingly, such a scenario also predicts an increase in $N_{\rm{eff}}$ with respect to the Standard Model value~\cite{Serpico:2004nm}. Typically, one expects $N_{\rm{eff}} \in [3.1,3.8]$ by combining the most recent CMB and Big Bang Nucleosynthesis data~\cite{Boehm:2013jpa,Boehm:2012gr,Nollett:2013pwa,Steigman:2013yua}. This is entirely compatible with the value of $N_{\rm{eff}} = 3.5 \pm 0.4$ obtained in Sec.~\ref{subsec:lss} when we impose our Lyman-$\alpha$ limit. As a result, we predict a rather higher value of $H_0 = 71 \pm 3~{\mathrm{km}}~{\mathrm{s}}^{-1}~{\mathrm{Mpc}}^{-1}$ (see Table~\ref{tab:params3}), in good agreement with direct measurements of the local Hubble parameter.

Finally, it is worth noting that in this toy model, one expects the (radiative) generation of small neutrino masses. Assuming ${\mathcal{O}}(1) \ \rm{MeV} \lesssim$ $m_N$ $\lesssim 10 \ \rm{MeV}$, one obtains neutrino masses in the range $0.01 \ \rm{eV} \lesssim$ $m_{\nu}$ $\lesssim 1 \ \rm{eV}$ provided that the coupling, $g$, satisfies~\cite{Boehm:2006mi}:
\begin{eqnarray}
g &\simeq& 10^{-3} \ \sqrt{\frac{m_N}{10 \ \rm{MeV}}} \ \left( \frac{\langle \sigma v \rangle}{3 \times 10^{-26}~{\rm cm}^3~{\rm s}^{-1}} \right)^{1/4} \nonumber \\
&\times& \sqrt{1+{(m_{\rm DM}/{m_N})}^2}~.
\label{grd}
\end{eqnarray}

In the case of a strong mass degeneracy between the DM particle and the mediator, Eq.~\eqref{grd} gives
\begin{equation}
g \simeq 10^{-3} \ \sqrt{\frac{m_N}{10 \ \rm{MeV}}} \ \left( \frac{\langle \sigma v \rangle}{3 \times 10^{-26}~{\rm cm}^3~{\rm s}^{-1}} \right)^{1/4}~,
\label{grd2}
\end{equation}
which is compatible with Eq.~\eqref{eq:g} for MeV DM.

In summary, for this specific realisation, we expect a cut-off in the matter power spectrum at the Lyman-$\alpha$ scale, a departure of $N_{\rm{eff}}$ from the Standard Model value,
$H_0 \sim 71 \pm 3~{\mathrm{km}}~{\mathrm{s}}^{-1}~{\mathrm{Mpc}}^{-1}$ and the generation of neutrino masses.

Our model assumes a strong mass degeneracy between the DM particle and the mediator, but this could be suggestive of an exact symmetry in the invisible sector (such as unbroken Supersymmetry, without any counterpart in the visible sector). The other requirement is particles in the MeV mass range. Such properties may be challenging to realise in a theoretical framework, yet the model building remains to be done.  

Expressions for the DM--neutrino elastic scattering cross section with a Dirac or Majorana DM candidate can be found in Ref.~\cite{Boehm:2003hm}. When there is a strong mass degeneracy, the cross section is expected to be constant and proportional to
\begin{equation}
\sigma_{\rm{DM}-\nu} \propto \frac{g^4}{m_{\rm{DM}}^2}~,
\end{equation}
as for the scalar case. The annihilation cross section is also given by a similar expression, so again, for specific values of $g$ (analogous to Eq.~\eqref{eq:g}), we expect a cut-off in the matter power spectrum at a relevant cosmological scale and simultaneously, the correct DM relic abundance.

In all the above scenarios, DM could potentially be produced by neutrinos in supernovae. However, here we do not consider a coupling to nucleons (DM is only coupled to neutrinos) and 
the cross section does not increase with temperature (it remains constant). Therefore, we do not expect a large impact on supernovae cooling, but this would need to be checked in a dedicated study.

\subsection{$T^2$-Dependent Cross Section}

If one relaxes the hypothesis of a strong mass degeneracy between the DM particle and the mediator, the DM--neutrino elastic scattering cross section becomes dominated by a term proportional to $T^2$ (independently of whether we consider a scalar or fermionic DM candidate). If we assume that neutrinos are Majorana particles, we obtain:
\begin{equation}
\sigma_{\rm{DM}-\nu}  \propto \frac{g^4}{\pi}  \ \frac{T^2 }{m_N^4 } \ + \ \mathcal{O}(T^3)~,
\label{eq:no_mass_degen}
\end{equation}
which leads to
\begin{equation}
\sigma_{\rm{DM}-\nu} \simeq 10^{-46} \ A \ \left(\frac{g}{0.1}\right)^4 \   \left(\frac{T}{T_0}\right)^{2}  \ \left(\frac{m_N}{\rm{MeV}}\right)^{-4} \ \rm{cm^2}~,
\label{sigmaelT2}
\end{equation} 
where $A$ is a numerical factor that depends on the exact nature of the DM particle and $T_0 \simeq 2.3 \ 10^{-4} \ \rm{eV} $ is the temperature of the Universe today.

Therefore, one expects a damping in the matter power spectrum at the Lyman-$\alpha$ scale if the DM mass is in the MeV range and $g\sim 0.1 \times (m_{N}/{\rm MeV})$.
For such a configuration, there could be, in addition, a resonance feature in the diffuse supernovae neutrino background~\cite{Farzan:2014gza}.

If neutrinos have only right-handed couplings and we do not impose a very strong degeneracy between $m_N$ and $m_{\rm{DM}}$, the cross section remains $T^2$-dependent. Its value would be 
of the same order as the Lyman-$\alpha$ bound provided that the DM mass is again in the MeV range and the mass splitting between the mediator and the DM particle is relatively small (about 10$\%$).

A $T^2$-dependent cross section is easier to achieve than a constant cross section described in Sec.~\ref{subsec:const} since it does not require the mediator and the DM particle to be mass degenerated. However, the observed DM abundance would be difficult to explain in the thermal case as the annihilation cross section would be too large for $g \gtrsim 0.1$ (although solutions exist e.g. asymmetric DM~\cite{Kaplan:2009ag}). One would also lose the relation with the neutrino masses. A similar conclusion is obtained for a DM candidate coupled to a new (weakly-coupled) gauge boson (see Ref.~\cite{Boehm:2003hm}).

\section{Conclusion}
\label{sec:conc}

In this paper, we have studied the effects of introducing an interaction between dark matter and neutrinos on the evolution of primordial matter fluctuations. Using cosmological data from Planck and the Lyman-$\alpha$ forest, we have obtained the following constraints: $\sigma_{\rm{DM}-\nu} \lesssim 10^{-33} \left(m_{\rm{DM}}/\rm{GeV}\right) \ \rm{cm^2}$ if the cross section is constant and $\sigma_{\rm{DM}-\nu,0} \lesssim 10^{-45} \left(m_{\rm{DM}}/\rm{GeV}\right) \ \rm{cm^2}$ if it scales as the temperature squared. Such results are importantly model-independent and can be applied to any theory beyond the Standard Model that predicts a coupling between dark matter and neutrinos.

In particular, we have seen that models involving thermal MeV DM and a constant scattering cross section can accommodate larger values of $N_{\rm eff}$ and $H_0$ with respect to $\Lambda$CDM, produce a cut-off in the matter power spectrum at the Lyman-$\alpha$ scale and at the same time, generate small neutrino masses.

\acknowledgements

The authors would like to thank Carlton Baugh, Claude Duhr, Silvia Pascoli, Martin Haehnelt, Christopher McCabe and Jascha Schewtschenko for useful discussions. We acknowledge the use of the publicly available numerical codes {\sc class} and {\sc Monte Python}, and the CMB data from the WMAP, Planck and SPT experiments. RJW and CB are supported by the STFC and the European Union FP7 ITN INVISIBLES.


\bibliography{DMNU.bib}


\end{document}